\documentclass[12pt,a4paper]{article}
\usepackage{graphicx}
\usepackage{amsmath}
\usepackage{amssymb}
\usepackage{bm}
\usepackage{multirow}
\usepackage{cases}
\usepackage{geometry}

\begin{document}

\title{Light transport behaviours in quasi-1D disordered waveguides composed of random photonic lattices}

\author{Yuchen Xu, Hao Zhang$^{*}$,Yujun Lin and Heyuan Zhu\\ \textit{\small Shanghai Ultra-precision Optical Manufacturing Engineering Center,} \\ \textit{\small Department of Optical Science and Engineering,} \\ \textit{\small Fudan University, Shanghai 200433, China}\\ \small $^*$zhangh@fudan.edu.cn}


\date{}
\maketitle
\begin{abstract}
We present a numerical study on the light transport properties which are modulated by the disorder strength in quasi-one-dimensional disordered waveguide which consists of periodically arranged scatterers with random dielectric constant. The transport mean free path is found to stay inversely proportional to the square of the relative fluctuation of the dielectric constant as in the 1D and 2D cases but with . The transport properties of light through a sample with a fixed size can be modulated from ballistic to localized regime as well, and a generalized scaling function is defined to determine the light transport status in such a sample. The calculation of the diffusion coefficient and the energy density profile of the most transmitted eigenchannel clearly exhibits the transition of transport behaviour from diffusion to localization. 
\end{abstract}

\section{Introduction}

Transport in random media at the mesoscopic scale has attracted much attention in recent decades. In the multiple scattering process, wave interference persists and leads to a series of extraordinary phenomena in contrast to common diffusion, such as Anderson localization and enhanced backscattering\cite{Anderson1958,Albada1985,Wolf1985}. Anderson localization predicts exponentially localized modes and a halt of diffusion when the disorder reaches a certain extent, while enhanced backscattering manifests itself as the precursor of Anderson localization showing an intensity enhancement factor of 2 in the opposite direction of the incident wave due to constructive interference\cite{Stoerzer2006}. About two decades after Anderson localization was predicted, a single-parameter scaling theory of localization was proposed\cite{Abrahams1979}. It defines a universal scaling function $\beta(g)=d\ln g/d\ln L$ which shows how the electronic dimensionless conductance $g$ of a random system decreases with the incremental system length $L$, without directly considering the disorder strength of the system.

Experimental observations of Anderson localization have been realized so far for matter waves\cite{Billy2008,Roati2008,Jendrzejewski2012}, elastic media\cite{Hu2008} and photons\cite{Stoerzer2006,Wiersma1997,Chabanov2000,Bliokh2006,Schwartz2007,Toninelli2008}. Among these, the experiments carried out by Schwartz et. al. adopted a real-time induction method to form two-dimensional photonic lattices with controlled disorder in a dielectric crystal, and showed the variation of the transport from ballistic to diffusive by increasing the disorder strength, and transverse localization is definitely observed when the disorder is strong enough\cite{Schwartz2007}. This reminds us that, one can take advantage of the disorder-modulation other than changing the length scale of the sample to investigate the transport properties in random media. 

For a fixed sample size $L$ in quasi-one-dimension (quasi-1D), the light propagation through a random media mainly depends on some length scales, especially the transport mean free path (TMFP) $l_{\mathrm{tr}}$. $l_{\mathrm{tr}}$ is the critical length scale over which the incident wave loses its initial direction. It is also the fundamental parameter measuring the disorder of a random medium which only depends on the disorder strength and the incident wavelength. The relation of $l_{\mathrm{tr}}$, $L$ and the localization length $\xi$ determines which transport regime does the system belong to. Therefore, the study on the transport mean free path should be in the first place when investigating the transport properties of random media. 

Investigation of transmission, based on the transmission matrix $t$ of the system, however, enables us to acquire comprehensive understanding on the transport properties of random media. The optical counterpart of the electronic dimensionless conductance $g$, i.e., the transmittance $T$, can be expressed as the sum of all the transmission eigenvalues as $T=\sum_{n=1}^{N} \tau_{n}$, where $\{\tau_n\}$ are the transmission eigenvalues lying within the range from 0 to 1(full transmission) and $N$ is the number of transverse modes. $\tau_n$ can be obtained from the singular value decomposition of the transmission matrix $t=\sum_{n=1}^{N} \textbf{u}_{n}\sqrt{\tau_n}\textbf{v}_{n}^{\dagger}$, with $\textbf{u}_{n}$ and $\textbf{v}_{n}$ the transmission eigenchannels composing the incoming and outgoing modes\cite{Mello1988}. In the ballistic regime, $T$ satisfies the macroscopic transport theory, while in the diffusive regime, $T\propto Nl_{\mathrm{tr}}/L$, and when $T$ is close to unity, the system is about to fall into localized regime, where the eigenchannel with the maximal transmission eigenvalue dominates. The transmittance $T$ thus gives an overall description of the transport properties. 

Since the sample length is finite, the diffusion within the sample reveals position-dependent characteristics\cite{Tiggelen2000,Tian2010}. For a finite sample, the variation of disorder may influence the local diffusion coefficient $D(x)$ since the diffusion related parameter $l_{\mathrm{tr}}$ is modulated. 

In this paper, we first investigate the relation between the TMFP $l_{\mathrm{tr}}$ and the variant disorder strength with the sample length $L$ fixed in a quasi-1D disordered waveguide using the exact Anderson disorder model . With $l_{\mathrm{tr}}$ determined, the wave transport properties including the transmittance $\left<T\right>$ and the diffusion coefficient $D(x)$ are calculated to show the transport regime transition with the increasing disorder and verify the feasibility of modulating the transport behaviour in a periodical lattice of scatterers with random dielectric constant. 

\section{Numerical Methods}
\begin{figure}[ht]
\centering
\includegraphics[width=10cm]{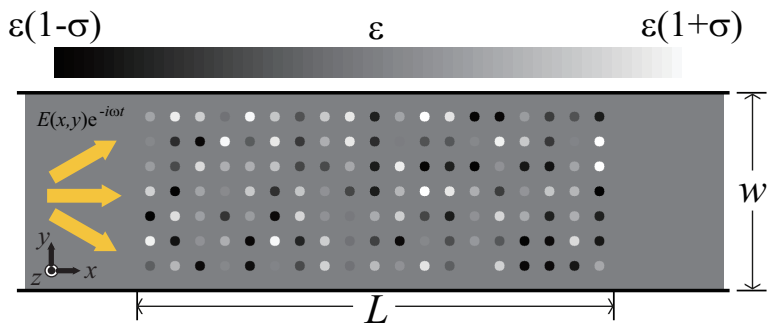}
\caption{Schematic of the Quasi-1D waveguide considered in the simulations.}
\label{Fig:q1d}
\end{figure}

Consider a quasi-one-dimensional (quasi-1D) disordered sample(locally 2D) of length $L$ and width $w$, with two identical semi-infinite free waveguides attached to its both ends, which are non-reflective due to the dielectric constant matching, and the transverse boundaries of the entire system is perfectly reflective, as shown in Fig.~\ref{Fig:q1d}. A monochromatic light wave $E(x,y)\exp(-i\omega t)$ of $z$-polarization propagating along the $x$-direction is governed by the Helmholtz equation 
\begin{equation}
\nabla^{2}E(x,y)+k^{2}[1+\mu(x,y)]E(x,y)=0, \label{helmholtz}
\end{equation}
where $k=\sqrt{\varepsilon}k_{0}$ is the wave vector, with $\varepsilon$ the dielectric constant of the free waveguide, $k_{0}=\omega/c$ the wave vector in vacuum, respectively, and $\mu(x,y)=\delta\varepsilon(x,y)/\varepsilon$ is the relative fluctuation of the dielectric constant, which induces the disorder of the sample. 

The quantized eigenstates of an empty waveguide are 
\begin{equation}
\varphi^{(\pm)}_{n}(x,y)=\frac{1}{\sqrt{k_{n}}}\chi_{n}(y)e^{\pm ik_{n}x}, \label{freestates}
\end{equation}
where $k_{n}$ is the longitudinal wave vector, $\chi_{n}(y)$ is the transverse wave function, the positive integer $n(1\leqslant n \leqslant N)$ is the index of the eigenchannel, where $N \propto kw$ is the total number of the eigenchannels. The factor $k_{n}^{-1/2}$ here ensures that each channel carries normalized flux. Under the perfect reflection boundary condition, the transverse wave function takes the form 
\begin{equation}
\chi_{n}(y)=\sqrt{\frac{2}{w}}\sin\left( \frac{n\pi y}{w} \right)
\end{equation}
and the corresponding longitudinal wave vector is 
\begin{equation}
k_{n}=\sqrt{k^2-\left( \frac{n\pi}{w} \right)^2}. 
\end{equation}
The transmission matrix $t$ can be calculated with the relation\cite{Fisher1981}
\begin{equation}\label{telement}
t_{ba}(x,x') = \sqrt{v_{b}v_{a}}\int_{0}^{w} \mathrm{d}y \int_{0}^{w} \mathrm{d}y' \chi_{b}^{*}(y)G^{\mathrm{r}}(x,y;x',y')\chi_{a}(y'). 
\end{equation}
Here $t_{ba}(x,x')$ is the element of $t(x,x')$(transmission matrix from one surface at $x'$ to another surface at $x$) which represents the complex field transmission amplitude from the incoming channel $a$ to the outgoing channel $b$. $G^{r}$ is the retarded Green's function. $v_{n}$ is the group velocity at the incident wavelength of the $n^{\mathrm{th}}$ channel. 

In the Anderson disorder model\cite{Anderson1958}, the waveguide is discretized into a square lattice with the coordinate discretization $x \rightarrow n_{x}d, y \rightarrow n_{y}d$, where $d$ is the lattice constant. The disordered region corresponds to $1 \leqslant n_{x} \leqslant N_{x}, 1 \leqslant n_{y} \leqslant N_{y}$, where $N_{x}=L/d, N_{y}=w/d$. In the simulations $k_{0}d$ is set to unity. In this basis, Eq.~\eqref{telement} can be written in a compact form as 
\begin{equation}
t^{n_{x},n'_{x}} = \frac{1}{d}V^{\dagger} X^{\dagger} G^{n_{x},n'_{x}} X V,  \label{tmatrix}
\end{equation}
where $V=\mathrm{diag}\{v_{n}^{1/2}\}$ is the diagonal matrix whose diagonal elements are the group velocities of the eigenchannels and $X=[\chi_{n_{y}n}]_{N_{y} \times N}$ is the matrix whose columns are the discretized transverse wave functions of the eigenchannels. $G^{n_{x},n'_{x}}$ is the entire Green's function connecting the slices indexed by $n_{x}$ and $n'_{x}$, which can be calculated using the recursive Green's function(RGF) method\cite{Mackinnon1985,Baranger1991}. Since advanced Green's function is not used here, the superscript ``r'' denoting retarded Green's function is omitted without ambiguity. To calculate the total transmission matrix through the whole disordered region, one just need to take $n_{x}=N_{x}+1$ and $n'_{x}=0$ in Eq.~\eqref{tmatrix}. 

The local diffusion coefficient $D(x)$ can be calculated with the first Fick's law 
\begin{equation}
\left< J(x) \right> = -D(x)\frac{\mathrm{d}\left< W(x) \right>}{\mathrm{d}x}, 
\end{equation}
where $J(x)$ is the energy flow, $W(x)$ is the local energy density, and $\left< \cdots \right>$ represents ensemble average. Since there is no absorption, the energy flow is conserved and $J(x)$ is replaced by a constant $J_{0}$. $W(x)$ can be calculated by 
\begin{equation}
W(x)=\sum_{n,n'}\left| \left[ t(x,0)\textbf{v}_{n} \right]_{n'} \right|^{2},
\end{equation}
where $t(x,0)\textbf{v}_{n}$ is the local field at $x$ induced by the $n^{\mathrm{th}}$ incoming channel. 
\section{Results and Discussions}
Since the transport mean free path $l_{\mathrm{tr}}$ depends on the the disorder strength of the sample and the incident wavelength, thus $l_{\mathrm{tr}}$ is determined by $\mu(x,y)$ when $k$ is fixed. By applying the Anderson disorder model, $\mu(x,y)$ is quantized to $\mu_i$ on the site $i$ in a square lattice. $\{\mu_{i}\}$ are independent and identically distributed random variables, therefore they satisfy 
\begin{equation}\label{corr}
\left<\mu_{i}\mu_{j}\right> = 
\begin{cases}
\frac{\sigma^{2}}{3}\delta_{ij}, &  \mu_{i} \sim U(-\sigma,\sigma)  \\
\sigma^{2}\delta_{ij}, &  \mu_{i} \sim N(0,\sigma) 
\end{cases}
\end{equation}
where $U(-\sigma,\sigma)$ stands for the uniform distribution and $N(0,\sigma)$ stands for the standard normal distribution, and for both distributions, obviously, $\left<\mu_{i}\right> = 0$. 

\begin{figure}[ht]
\centering
\includegraphics[width=10cm]{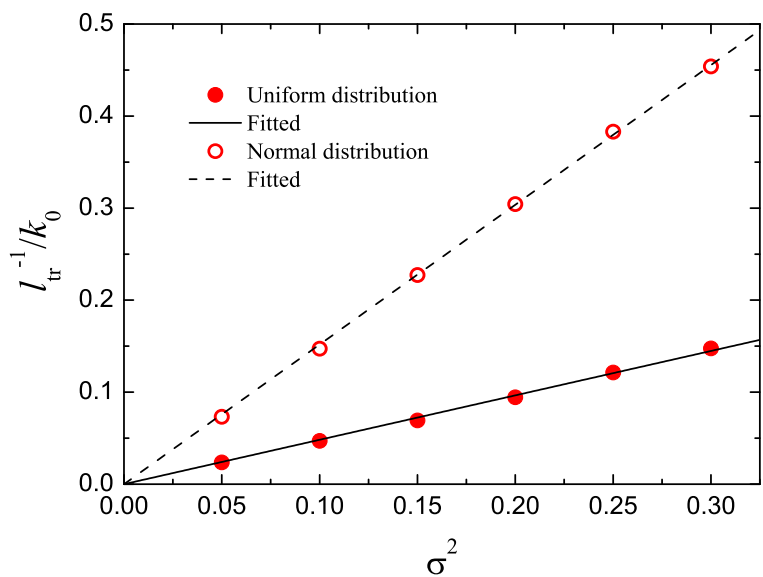}
\caption{Transport mean free path of a disordered sample in a waveguide with $N=5$ channels ($w=9/k_0$) and the environment dielectric constant $\varepsilon=2.25$, for relative dielectric constant fluctuation of a uniform distribution within [$-\sigma,\sigma$] (solid circles) and a normal distribution with a standard deviation $\sigma$ (hollow circles), which are fitted proportionally (solid line for the former and dashed line for the latter).}
\label{Fig:mfp}
\end{figure}

In the localized regime, the ensemble average of the logarithmic transmittance is proportional to the sample length and inversely proportional to the localization length $\xi$, i.e., $\left<\mathrm{ln}T\right> =-2L/\xi$. Based on the Anderson disorder model, and for both cases of distribution of $\mu_i$, $\left<\mathrm{ln}T\right>$ is calculated at 8 different sample lengths for each value of $\sigma$, where the average is performed over a sub-ensemble of 2000 disordered configurations. By fitting the slope of $\left<\mathrm{ln}T\right>$, we obtain the localization lengths, and the transport mean free paths can be extracted from the Thouless relation $\xi=(\pi/2)Nl_{\mathrm{tr}}$, where the channel number $N$ equals to 5. The calculated TMFPs are shown in Fig.~\ref{Fig:mfp} as solid circles for the uniform distribution and empty circles for the normal distribution of $\mu_i$. 

In the limit case $\sigma=0$, the sample is actually a perfect crystal with an infinite transport mean free path. According to the Green's function theory for quantum transport\cite{Akkermans2007b}, the mean free path is inversely proportional to the imaginary part of the self-energy $\mathrm{Im}\Sigma^{R}(k)$, where the superscript $R$ represents the retarded Green's function. The self-energy is proportional to the correlation function $\left<\mu_i^2\right>$, thus the mean free path is inversely proportional to the correlation function, i.e., $l_{\mathrm{tr}}^{-1} \propto \left<\mu_{i}^{2}\right>$. The mean free paths for the uniform distribution and the normal distribution are linearly fitted separately, and the fitted slopes are 0.4825 and 1.518, respectively.  The two slopes shows a difference factor $\approx 3$ which results from the the correlation function of the variation in the dielectric constant for different distributions as indicated by Eq.~\eqref{corr}. $l_{\mathrm{tr}}$ is inversely proportional to $\sigma^{2}$ as predicted by theory, i.e., $l_{\mathrm{tr}}=l_{0}\sigma^{-2}$, where $l_{0}$ is the TMFP for $\sigma=1$, which equals to $24/(\pi\varepsilon k)\simeq 2.264/k_{0}$ for 2D systems and $12/(\pi\sqrt{\varepsilon}k)\simeq 1.698/k_{0}$ for 1D systems with $\mu_{i}$ uniformly distributed\cite{Berkovits1994}. However, our simulation obtains $l_{0}\simeq 2.073/k_{0}$, which is between the values for 2D and 1D cases. This implies that light loses its initial direction faster in a quasi-1D (locally 2D) disordered waveguide than in a 2D one due to the transverse confinement of the perfectly reflecting boundaries. 

Since the light localization length is very large in samples with very weak disorder, it will take much time to calculate the TMFP in a long sample. The linear relation of $l_{\mathrm{tr}}$ and $\sigma^{-2}$ for optical disordered systems obtained herein can improve the computation efficiency especially for weakly disordered system by considering a relatively short sample with strong disorder. 


\begin{figure}[ht]
\centering
\includegraphics[width=12cm]{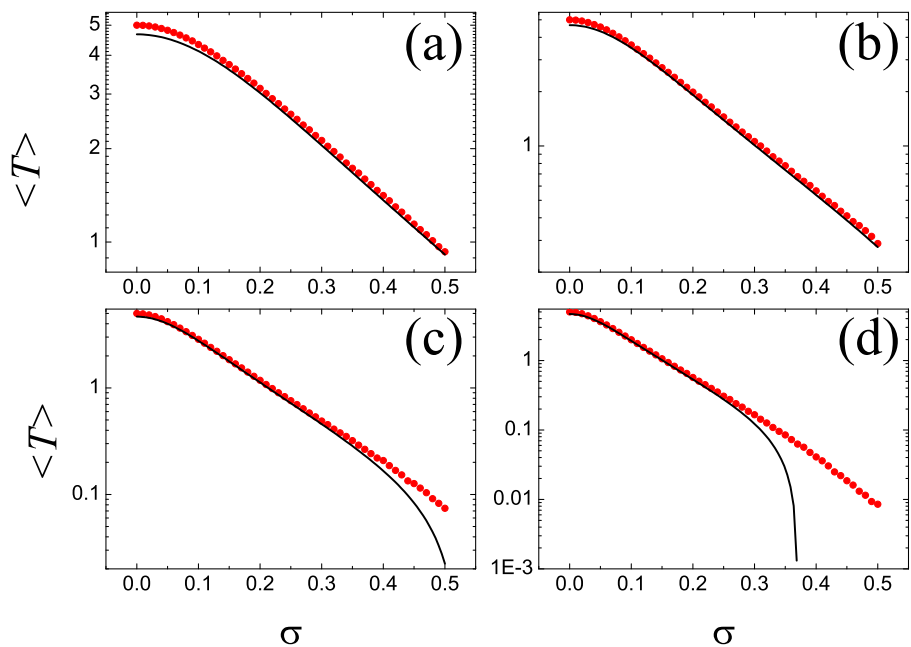}
\caption{Dependence of transmittance $\left<T\right>$ on the disorder strength $\sigma$ within the range [0,0.5] for four different sample lengths (a) $L=40/k_{0}$; (b) $L=100/k_{0}$; (c) $L=200/k_{0}$; (d) $L=400/k_{0}$. The environmental dielectric constant is $\varepsilon=2.25$ and the channel number is $N=5$. }
\label{Fig:transmittance}
\end{figure}

With the TMFP determined, the average transmittance $\left<T\right>$ as a function of the disorder strength $\sigma$ is calculated and shown in Fig.~\ref{Fig:transmittance} for 4 different sample lengths (only consider the uniformly distributed $\mu_{i}$ from now on). The simulated results are compared with the consistent results of several theoretical methods which expand $\left<T\right>$ as\cite{Mello1991,Mirlin2000,Payne2010} 
\begin{equation}\label{gexpand}
\left<T\right> \simeq g_{0}-\frac{1}{3}+\frac{1}{45g_{0}}+\frac{2}{945g_{0}^{2}}+\mathcal{O}\left(\frac{1}{g_{0}^{3}}\right), 
\end{equation}
where 
\begin{equation}\label{bareg}
g_{0}=\frac{N}{1+2L/\pi l_{\mathrm{tr}}}
\end{equation}
is the bare conductance. Eq.~\eqref{gexpand} is only valid in the diffusive transport regime, i.e., $l_{\mathrm{tr}} < L \ll \xi$, and in Eq.~\eqref{bareg} the extrapolation length $z_{0}$ induced by the internal reflection at the two end of the disordered waveguide is taken into account and its value is $\pi l_{\mathrm{tr}}/4$. It should be emphasized that, with $l_{0}$ determined by fitting the linear relation between $l_{\mathrm{tr}}^{-1}$ and $\sigma^{2}$ before, there is no adjustable parameter in the fitting process herein. As shown in Fig.~\ref{Fig:transmittance}, the theoretical results fit well with our simulation results in the diffusive regime, which in turn verifies our results of the TMFP. 

When $\sigma$ is close to zero, i.e., in the ballistic and sub-diffusive regimes, the deviation of Eq.~\eqref{gexpand} from the simulation results is most obvious and mainly comes from the second term $-1/3$ (not negligible compared to the relatively small channel number $N$ considered here), which is a signal effect of the weak-localization correlation\cite{Mello1991} and is absent when scattering is very weak. In this case, the leading term of Eq.~\eqref{gexpand}, i.e., the bare conductance $g_{0}$ is enough to describe the transmission behaviour. 

For $L=40/k_{0}$ (Fig.~\ref{Fig:transmittance}(a)) and $L=100/k_{0}$ (Fig.~\ref{Fig:transmittance}(b)), the wave transport never really enters the localized regime ($L \gg \xi$). For $L=200/k_{0}$ (Fig.~\ref{Fig:transmittance}(c)), the transport can cover the three regimes with $\sigma$ tuned from $0$ to $0.5$, and for $L=400/k_{0}$, it enters the localized regime at a smaller value of $\sigma$ about $0.25$. This value corresponds to a maximal fluctuation of the refractive index of about $0.2$, which can be easily realized by doping randomly distributed scatterers into common optical materials and is also promising in systems of periodically arranged scatterers with random refractive index (close to our simulation model).

\begin{figure}[ht]
\centering
\includegraphics[width=10cm]{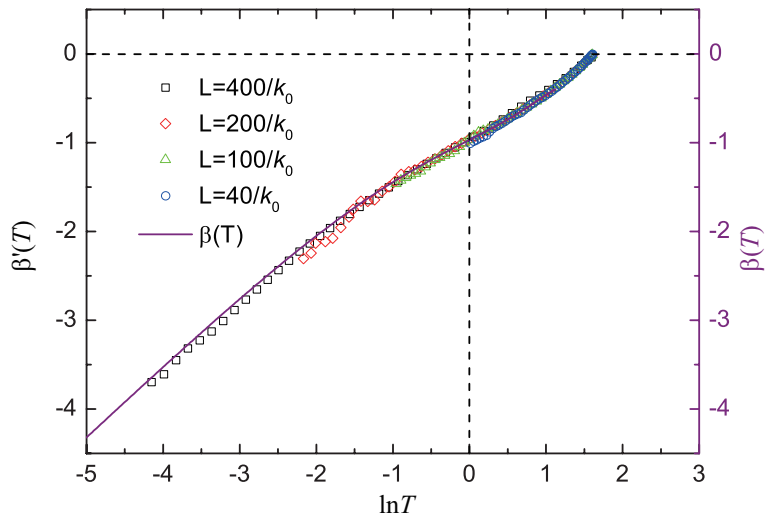}
\caption{The generalized scaling function $\beta'(T)$ calculated for samples with channel number $N=5$ and dielectric constant $\varepsilon=2.25$. The blue circles, green triangles, red diamonds and black squares indicate sample length $40/k_0, 100/k_0$, $200/k_0$ and $400/k_0$, respectively. As a comparison, the classical scaling function is plotted in line.}
\label{Fig:scaling}
\end{figure}

The classical scaling function $\beta(T)$ defined in Ref. \cite{Abrahams1979} is obtained by taking the derivative $\mathrm{d}\ln T/\mathrm{d}\ln L$ (the angular brackets $\left<\cdots\right>$ representing the ensemble average is dropped for simplicity), but with the disorder strength fixed at $\sigma=0.5$ and the channel number fixed at $N=5$, which is plotted in line in Fig.~\ref{Fig:scaling}. When $L$ is fixed,  it is straightforward to generalize the scaling function by including the variant $\sigma$ as 
\begin{equation}
\beta'(T)=\frac{\mathrm{d}\ln T}{\mathrm{d}\ln L'}, 
\end{equation}
where $L'=L\sigma^{2}$ is the effective sample length. When the transport is ballistic, $T$ increases to $N$ with a gradually vanishing increasing rate when $\sigma$ decreases to 0, which results that $\beta'(T)$ increases to 0 at $\ln T=\ln N$, as can be seen in Fig.~\ref{Fig:scaling}. The vanishing point of $\beta'(T)$ moves right when $N$ increases, which means that the scaling behaviour is influenced by the width of the sample as well. 

In the localized regime, $T$ is of the form $T=T_a \exp (-\gamma L')$, where $T_a$ is some critical transmittance of order unity and $\gamma \propto 1/Nl_{0}$. Thus $\beta'(T)$ for $T \rightarrow 0$ can be derived as 
\begin{equation}
\lim_{T \rightarrow 0}\beta'(T)=\ln(T/T_a), 
\end{equation}
as is shown in Fig.~\ref{Fig:scaling} for $\ln T \lesssim -2$. $\beta'(T)$ obtained from simulations for 4 different sample lengths is plotted in Fig.~\ref{Fig:scaling} with discrete markers. The coincidence of the line and the markers shows that it is equivalent to describe the transport status using $\beta'(T)$ and $\beta(T)$ and $L'$ determines the transmission for a fixed $N$, which confirms the feasibility of the generalization of the scaling function. 

\begin{figure}[ht]
\centering
\includegraphics[width=10cm]{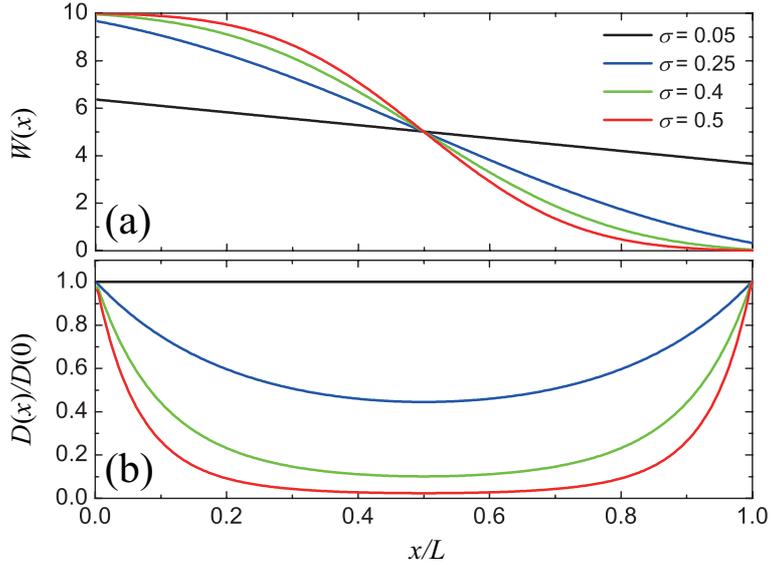}
\caption{Energy density profiles are shown in (a) for 4 different $\sigma=0.05,0.25,0.4$ and $0.5$, with the sample length fixed at $L=400/k_0$; Local diffusion coefficients normalized by the diffusion constant on the incident surface of sample, which are obtained from the energy densities are shown in (b). }
\label{Fig:diffcoeff}
\end{figure}

The distribution of light energy density and the local diffusion coefficient can describe the light transport behaviour through random media of different disorder strength in detail. The calculated energy density profiles $W(x)$ and the local diffusion coefficients $D(x)$ (normalized by $D(0)$) for different disorder strength are shown in Fig.~\ref{Fig:diffcoeff}(a) and (b), respectively, with the sample length fixed at $L=400/k_0$, and 20,000 realizations taken to perform the ensemble average. 

As shown in Fig.~\ref{Fig:diffcoeff}(a), the ensemble-averaged energy density monotonically decreases from the incident boundary to the output boundary, despite the value of $\sigma$. The backscattering of light leads to the decrease of energy density along the transport direction. Open boundaries of the sample cause energy leakage from them and the inhomogeneity of light interference. When near the boundaries, energy leaks out more easily and the interference is weaker. The descending rate keeps invariant when the disorder is very weak, e.g., for $\sigma=0.05$, since the scattering is very weak and the energy scattered out from the boundaries is negligible. Thus the case for $\sigma=0.05$ is a classical diffusive process with a constant diffusion coefficient, which is confirmed in Fig.~\ref{Fig:diffcoeff}(b). When $\sigma$ increases, the interference inhomogeneity and the energy leakage is strengthened and becomes significant, which make the descending rate of the energy density become position-dependent. 

Diffusion coefficients are obtained from the negative inverse derivative of the energy densities, as shown in Fig.~\ref{Fig:diffcoeff}(b). When the disorder is very weak, the diffusion coefficient is position-independent, while as the disorder strength increases, position-dependence emerges due to the facts that the wave interference is inhomogeneous and the returning probability becomes larger when gradually leaving the surface deepening into the sample. The position-dependence of the diffusion coefficient is a signal of localization\cite{Tiggelen2000,Tian2010}. 

\begin{figure}[ht]
\centering
\includegraphics[width=10cm]{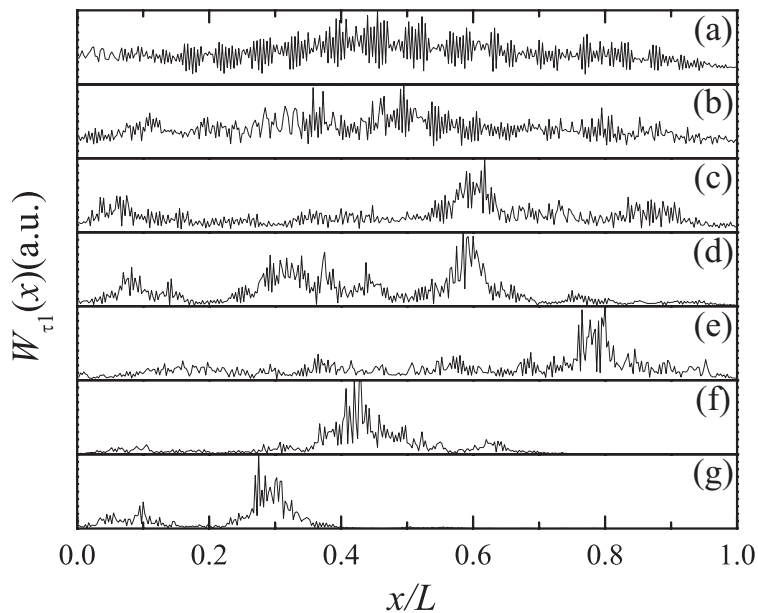}
\caption{Normalized energy density profiles $W_{\tau_{1}}(x)$ of the eigenchannel with the maximal eigenvalue $\tau_{1}$ for $\sigma^{2}=0.01,0.025,0.05,0.1,0.15,0.2,0.25$ (from top to bottom). }
\label{Fig:edpch1}
\end{figure}

The energy density profile $W_{\tau_{1}}(x)$ (for single disorder configuration) of the eigenchannel with the maximal transmission eigenvalue $\tau_{1}$ also qualitatively but directly exhibits the evolution of the transport from diffusion to localization, as shown in Fig.~\ref{Fig:edpch1}. For $\sigma^{2}=0.01$ (the uppermost curve), the energy density is extended through the whole disordered region, while for $\sigma^{2}=0.25$ (the nethermost curve), the energy density is localized in a short range along the $x-$direction. Fig.~\ref{Fig:edpch1} thus gives an intuitive view of the process of turning a diffusive disordered sample into a localized one by increasing $\sigma$. 


\section{Conclusions}
We have performed detailed investigations to show how the disorder influences the light transport properties in a quasi-1D random system. We calculated the disorder-modulated transport mean free path and found a scaling relation with the disorder strength parameter $\sigma$. With this relation the transport properties of light are considered, and we found that when the sample length is fixed, the disorder modulation enables the system to cover all the transport regimes. We also discussed the influence of the disorder modulation on the diffusion coefficient for samples with finite sizes, and confirmed that the diffusion is influenced by the strong inhomogeneous interference, especially in the localized regime. 

This work is supported by the National Natural Science Foundation of China under Grant No. 11374063, and 973 Program(No. 2013CAB01505).


\begin{thebibliography}{10}

\bibitem{Anderson1958}
P.~W. Anderson.
\newblock Absence of diffusion in certain random lattices.
\newblock {\em Phys. Rev.}, 109:1492--1505, Mar 1958.

\bibitem{Albada1985}
Meint P.~Van Albada and Ad~Lagendijk.
\newblock Observation of weak localization of light in a random medium.
\newblock {\em Phys. Rev. Lett.}, 55:2692--2695, Dec 1985.

\bibitem{Wolf1985}
Pierre-Etienne Wolf and Georg Maret.
\newblock Weak localization and coherent backscattering of photons in
  disordered media.
\newblock {\em Phys. Rev. Lett.}, 55:2696--2699, Dec 1985.

\bibitem{Stoerzer2006}
Martin St\"orzer, Peter Gross, Christof~M. Aegerter, and Georg Maret.
\newblock Observation of the critical regime near anderson localization of
  light.
\newblock {\em Phys. Rev. Lett.}, 96:063904, Feb 2006.

\bibitem{Abrahams1979}
E.~Abrahams, P.~W. Anderson, D.~C. Licciardello, and T.~V. Ramakrishnan.
\newblock Scaling theory of localization: Absence of quantum diffusion in two
  dimensions.
\newblock {\em Phys. Rev. Lett.}, 42:673--676, Mar 1979.

\bibitem{Billy2008}
Juliette Billy, Vincent Josse, Zhanchun Zuo, Alain Bernard, Ben Hambrecht,
  Pierre Lugan, David Clement, Laurent Sanchez-Palencia, Philippe Bouyer, and
  Alain Aspect.
\newblock Direct observation of anderson localization of matter waves in a
  controlled disorder.
\newblock {\em Nature}, 453(7197):891--894, June 2008.

\bibitem{Roati2008}
Giacomo Roati, Chiara D'Errico, Leonardo Fallani, Marco Fattori, Chiara Fort,
  Matteo Zaccanti, Giovanni Modugno, Michele Modugno, and Massimo Inguscio.
\newblock Anderson localization of a non-interacting bose-einstein condensate.
\newblock {\em Nature}, 453(7197):895--898, June 2008.

\bibitem{Jendrzejewski2012}
F.~Jendrzejewski, A.~Bernard, K.~Muller, P.~Cheinet, V.~Josse, M.~Piraud,
  L.~Pezze, L.~Sanchez-Palencia, A.~Aspect, and P.~Bouyer.
\newblock Three-dimensional localization of ultracold atoms in an optical
  disordered potential.
\newblock {\em Nat Phys}, 8(5):398--403, May 2012.

\bibitem{Hu2008}
Hefei Hu, A.~Strybulevych, J.~H. Page, S.~E. Skipetrov, and B.~A. van Tiggelen.
\newblock Localization of ultrasound in a three-dimensional elastic network.
\newblock {\em Nat Phys}, 4(12):945--948, December 2008.

\bibitem{Wiersma1997}
Diederik~S. Wiersma, Paolo Bartolini, Ad~Lagendijk, and Roberto Righini.
\newblock Localization of light in a disordered medium.
\newblock {\em Nature}, 390(6661):671--673, December 1997.

\bibitem{Chabanov2000}
A.~A. Chabanov, M.~Stoytchev, and A.~Z. Genack.
\newblock Statistical signatures of photon localization.
\newblock {\em Nature}, 404(6780):850--853, April 2000.

\bibitem{Bliokh2006}
K.~Yu. Bliokh, Yu.~P. Bliokh, V.~Freilikher, A.~Z. Genack, B.~Hu, and
  P.~Sebbah.
\newblock Localized modes in open one-dimensional dissipative random systems.
\newblock {\em Phys. Rev. Lett.}, 97:243904, Dec 2006.

\bibitem{Schwartz2007}
Tal Schwartz, Guy Bartal, Shmuel Fishman, and Mordechai Segev.
\newblock Transport and anderson localization in disordered two-dimensional
  photonic lattices.
\newblock {\em Nature}, 446(7131):52--55, March 2007.

\bibitem{Toninelli2008}
Costanza Toninelli, Evangellos Vekris, Geoffrey~A. Ozin, Sajeev John, and
  Diederik~S. Wiersma.
\newblock Exceptional reduction of the diffusion constant in partially
  disordered photonic crystals.
\newblock {\em Phys. Rev. Lett.}, 101:123901, Sep 2008.

\bibitem{Mello1988}
PA~Mello, P~Pereyra, and N~Kumar.
\newblock Macroscopic approach to multichannel disordered conductors.
\newblock {\em Ann. Phys.}, 181(2):290--317, 1988.

\bibitem{Tiggelen2000}
B.~A. van Tiggelen, A.~Lagendijk, and D.~S. Wiersma.
\newblock Reflection and transmission of waves near the localization threshold.
\newblock {\em Phys. Rev. Lett.}, 84:4333--4336, May 2000.

\bibitem{Tian2010}
Chu-Shun Tian, Sai-Kit Cheung, and Zhao-Qing Zhang.
\newblock Local diffusion theory for localized waves in open media.
\newblock {\em Phys. Rev. Lett.}, 105:263905, Dec 2010.

\bibitem{Fisher1981}
Daniel~S. Fisher and Patrick~A. Lee.
\newblock Relation between conductivity and transmission matrix.
\newblock {\em Phys. Rev. B}, 23:6851--6854, Jun 1981.

\bibitem{Mackinnon1985}
A.~MacKinnon.
\newblock The calculation of transport properties and density of states of
  disordered solids.
\newblock {\em Z. Phys. B}, 59(4):385--390, 1985.

\bibitem{Baranger1991}
Harold~U. Baranger, David~P. DiVincenzo, Rodolfo~A. Jalabert, and A.~Douglas
  Stone.
\newblock Classical and quantum ballistic-transport anomalies in
  microjunctions.
\newblock {\em Phys. Rev. B}, 44:10637--10675, Nov 1991.

\bibitem{Akkermans2007b}
Eric Akkermans and Gilles Montambaux.
\newblock {\em Mesoscopic Physics of Electrons and Photons}.
\newblock Cambridge University Press, 2007.
\newblock Cambridge Books Online.

\bibitem{Berkovits1994}
Richard Berkovits and Shechao Feng.
\newblock Correlations in coherent multiple scattering.
\newblock {\em Physics Reports}, 238(3):135--172, March 1994.

\bibitem{Mello1991}
Pier~A. Mello and A.~Douglas Stone.
\newblock Maximum-entropy model for quantum-mechanical interference effects in
  metallic conductors.
\newblock {\em Phys. Rev. B}, 44(8):3559--3576, August 1991.

\bibitem{Mirlin2000}
Alexander~D. Mirlin.
\newblock Statistics of energy levels and eigenfunctions in disordered systems.
\newblock {\em Physics Reports}, 326(5鈥?6):259--382, March 2000.

\bibitem{Payne2010}
Ben Payne, Alexey Yamilov, and Sergey~E. Skipetrov.
\newblock Anderson localization as position-dependent diffusion in disordered
  waveguides.
\newblock {\em Phys. Rev. B}, 82(2):024205--, July 2010.

\end{thebibliography}
\end{document}